\documentclass[runningheads]{llncs}
\usepackage{graphicx}
\usepackage{amsfonts}
\usepackage{amssymb}
\usepackage{siunitx}

\begin{document}
\title{A Deep Learning Framework for Classification of \textsl{in vitro} Multi-Electrode Array Recordings}
\titlerunning{Deep Learning Classification for \textsl{in vitro} MEA}
%
\author{Yun Zhao\inst{1} \and
Elmer Guzman\inst{2,3} \and
Morgane Audouard\inst{2,3} \and
Zhuowei Cheng\inst{1} \and
Paul K. Hansma\inst{2} \and
Kenneth S. Kosik\inst{2,3} \and
Linda Petzold\inst{1}}

\authorrunning{Y. Zhao et al.}
%
\institute{Department of Computer Science, University of California, Santa Barbara 
\and
 Neuroscience Research Institute, University of California, Santa Barbara
\and 
 Department of Molecular Cellular and Developmental Biology, University of California, Santa Barbara\\
\email{yunzhao@cs.ucsb.edu}\\
}
\maketitle              
\begin{abstract}
Multi-Electrode Arrays (MEAs) have been widely used to record neuronal activities, which could be used in the diagnosis of gene defects and drug effects. In this paper, we address the problem of classifying \textsl{in vitro} MEA recordings of mouse and human neuronal cultures from different genotypes, where there is no easy way to directly utilize raw sequences as inputs to train an end-to-end classification model. While carefully extracting some features by hand could partially solve the problem, this approach suffers from obvious drawbacks such as difficulty of generalizing. We propose a deep learning framework to address this challenge. Our approach correctly classifies neuronal culture data prepared from two different genotypes --- a mouse Knockout of the delta-catenin gene and human induced Pluripotent Stem Cell-derived neurons from Williams syndrome. By splitting the long recordings into short slices for training, and applying Consensus Prediction during testing, our deep learning approach improves the prediction accuracy by 16.69\% compared with feature based Logistic Regression for mouse MEA recordings. We further achieve an accuracy of 95.91\% using Consensus Prediction in one subset of mouse MEA recording data, which were all recorded at six days \textsl{in vitro}. As high-density MEA recordings become more widely available, this approach could be generalized for classification of neurons carrying different mutations and classification of drug responses.

\end{abstract}
%

%
\section{Introduction}
Deep learning models have achieved remarkable success in computer vision~\cite{Imagenet}, speech recognition~\cite{Speech}, natural language processing~\cite{NLP} and the game of Go~\cite{Alphago}. Recently there has been increasing interest in using deep learning in end-to-end neuroscience data analysis~\cite{EEG1,EEG2,CNN_Neuron}. Inspired by biology, deep learning models share many common properties with neuron functions. Deep learning models enable the extraction of information from action potential recordings of neuron activity, playing a vital role in several important neuron-based research and application areas~\cite{Application}.

Convolutional neural networks (CNN) can learn local patterns in data by using convolution filters as their key components~\cite{CNN}. Originally developed for computer vision, CNN models have recently been shown to be effective for neuroscience data analysis. Deep learning has recently been used to identify abnormal EEG signals~\cite{EEG2}. In~\cite{EEG1}, researchers designed an end-to-end EEG decoding for movement-related information using deep CNNs. With the latest development in fabrication of MEAs, a CNN was used to classify different neuronal cell types using simulated in-vivo extracellular recordings~\cite{Simul1}. However, most of the work in this area has focused on simulated data~\cite{Simul1,Simul2} since the experimental \textsl{in vitro} recordings are too noisy and there are not sufficient training samples for deep learning models. Researchers have also manually extracted features for deep learning training~\cite{Simul1,Simul2}. However, this does not fully exploit the deep learning model's ability of end-to-end learning, which learns from the raw data without any prior feature selection.

MEAs with advanced neural probes have been widely utilized to measure neuronal activity by recording local field potential~\cite{MEA}. Since the same units are measured on multiple recording sites, MEA recordings provide rich spatial information, which could be used to help diagnose diseases and genetic abnormalities. Our objective in this work has been to develop a deep learning framework which can distinguish MEA recordings of different genotypes. For example, delta-catenin is a crucial brain-specific protein of the adherens junction complex that localizes to the postsynaptic and dendritic compartments. It is enriched in dendrites and can be localized to the post-synaptic compartment. Recent studies indicate that delta-catenin is required for the maintenance of neural structure and function in the mature cortex~\cite{LossD,RequiredD,JunctionD}. Williams syndrome (WS) is a neurodevelopmental disorder caused by a genomic deletion of about 28 genes~\cite{Williams,Williams1}. As a result of this hemideletion, the subjects display a characteristic phenotype with mild to moderate intellectual disability as well as behavioral features such as an outgoing personality and conserved communication skills. Studying those genes is of particular interest in order to decipher the social behaviors in humans~\cite{Williams2}.


In the present work, we propose an end-to-end CNN architecture to classify \textsl{in vitro} MEA recordings with different genotypes. We test our framework on mouse recordings to classify Wild Type and delta-catenin Knockout. We also attempt to classify human derived induced Pluripotent Stem Cell (iPSC) neuron cultures from Williams syndrome versus Control cultures. We split the long recordings into smaller slices for training to provide more training samples, and then apply Consensus Prediction during testing.

The key contributions of this paper include:

1) We propose a CNN based model to classify the genotype of \textsl{in vitro} MEA recordings, which outperforms Logistic Regression by 16.69\%. To the best of our knowledge, this is the first paper using deep leaning to classify \textsl{in vitro} MEA recordings.

2) We split the long recordings into smaller slices for training, which not only eases the burden on GPU memory but also provides many training samples for deep learning models.

3) We define Consensus Prediction as the majority voting result of the sampled short slices for testing, since not all of the short slices can be expected to contain enough useful information. We achieve an accuracy of 95.91\% using Consensus Prediction in one subset of MEA recording data, which were all recorded at 6 days \textsl{in vitro} (DIV). 

The rest of this paper is organized as follows. Section~\ref{Data} describes how our MEAs are recorded and introduces the classification problem. We delineate the deep learning model in Section~\ref{Model} and describe the experimental setup in Section~\ref{Setup}. Evaluation and discussion are provided in Sections~\ref{Evaluation} and~\ref{Discussion}, respectively.

\section{Data Collection and Classification}\label{Data}
\subsection{Mouse neuron culture}

Commercial MEAs (MultiChannel Systems) were sterilized with UV irradiation for $>$ 30 minutes, incubated with poly-L-lysine(0.1 mg/ml) solution for at least one hour at $37 ^{\circ}C$, rinsed several times with sterile deionized water and allowed to dry before cell plating. Wild-type mice were in a C57BL/6 background and littermate controls were obtained by breeding heterozygote male and female delta-catenin transgenic mice. For the delta-catenin transgenic mice, a targeted mutation in the delta-catenin gene is located within axon 9 of the delta-catenin locus and consists of a GFP reporter fused to a PGK-hyygro-pA cassette followed by a stop condon, which results in the prevention of transcription of the rest of the delta-catenin gene. Mouse pups were decapitated at P0 or P1, the brains were removed from the skulls and the hippocampi were dissected from the brain followed by manual dissociation and plating of 250,000 cells in the MEA chamber~\cite{Cut}. After one week, cultures were treated with 200 $\mu$M glutamate to kill any remaining neurons, followed by a new batch of cells added at the same density as before. Cultures were grown in a tissue culture incubator ($37 ^{\circ}C$, 5\% $CO_{2}$), in a medium made with Minimum Essential Media with 2 mM Glutamax (Life Technologies), 5\% heat-inactivated fetal calf serum (Life Technologies), 1 ml/L of Mito+ Serum Extender (BD Bioscience) and supplemented with glucose to an added concentration of 21 mM. All animals were treated in accordance with University of California and NIH policies on animal care and use.

\subsection{Culture of iPSCs neurons}
iPSCs were cultured in mTeSR1 media (Stem Cell Technologies) and routinely passaged with ReleSR (Stem Cell Technologies). The cells were subsequently infected with TetO-hNgn2-UBC-puro (plasmid from Addgene \# 61474) and rtTA (plasmid from Addgene \# 20342) lentiviruses. Briefly, the cells were passaged as single cells into 4 wells with accutase (Life Technologies) and Y-27632 dihydrochloride (Tocris) at a final concentration of 10 M. On day 2 the cells were infected with hNgn2 in fresh mTeSR1 media. On day 3, the infected iPSCs were selected by adding puromycin at 2 ug/ml for a 2 day period. The cells were infected with rtTA virus on day 5 and incubated overnight. The neurons were differentiated by adding doxycycline at a final concentration of 2 ug/ml. Two days after addition of doxycycline, the neurons were replated on poly-l-lysine coated MEAs at a density of 180,000 cells concentrated in a 15 ul droplet. iPSCs-derived neurons were cocultured with mouse primary astrocytes in BrainPhys complete medium (Stem Cell Technologies). Doxycycline was kept in the media for 14 days total.

\subsection{Electrophysiology}

\begin{figure}
\centering
\includegraphics[width=0.6\textwidth]{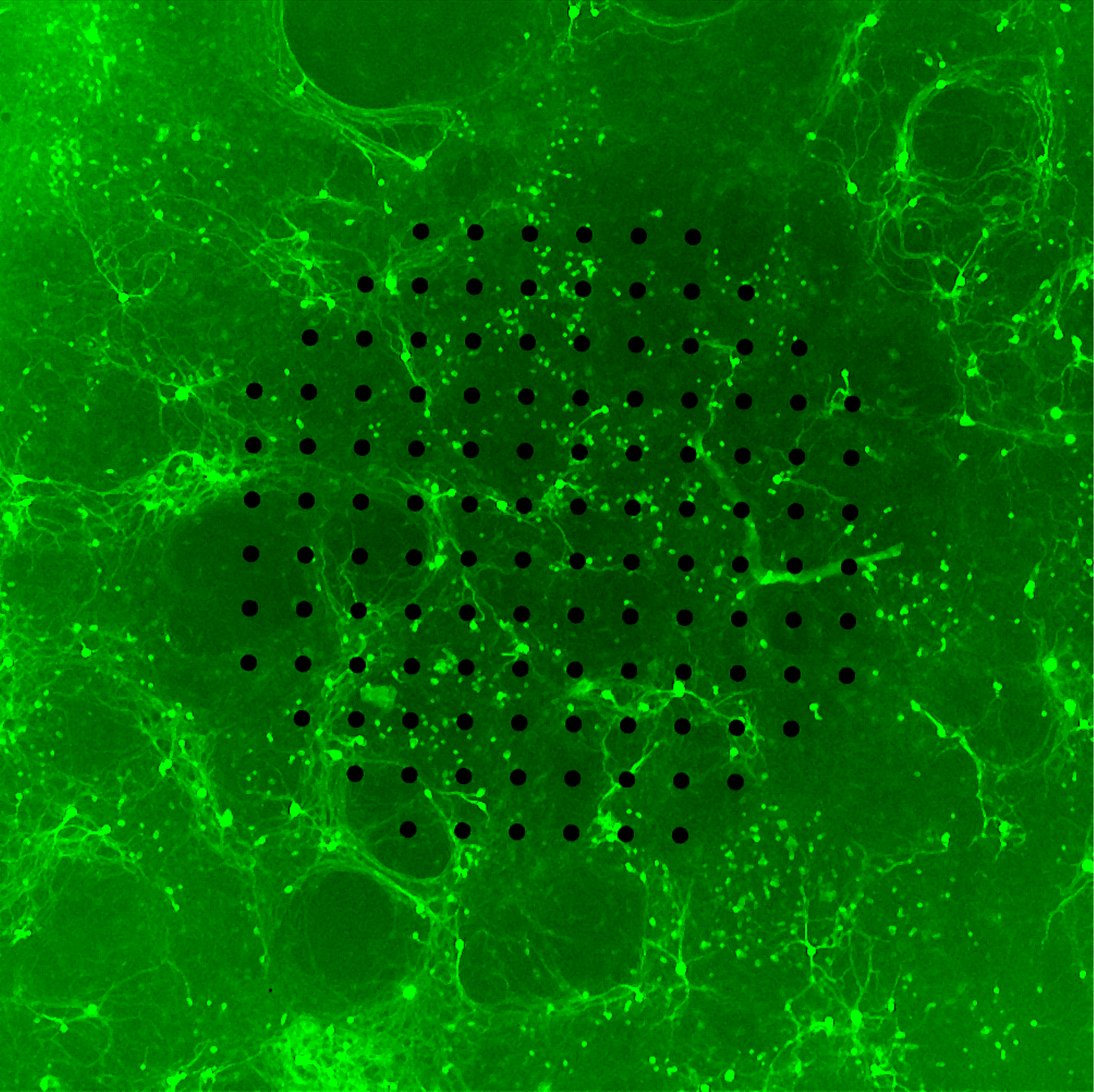}
\caption{Neural networks were grown on arrays of 120 electrodes. The purpose of this research was to determine whether neural cultures derived from genetically different neurons could be distinguished by analysis of their electrical activity.} \label{fig_MEA}
\end{figure}

We used 120 electrode MEAs (120MEA100/30iR-ITO arrays; MultiChannel Systems) for recording as is shown in Fig.~\ref{fig_MEA}. All recordings were performed in cell culture medium so as to minimally disturb the neurons. The osmolality of the culture medium was adjusted to 320 mosmol. Recordings were performed using MultiChannel Systems MEA 2100 acquisition system. Data were sampled at 20 kHz. Recordings were performed at $30 ^{\circ}C$. All recordings were performed on neurons at 2-30 DIV. Data recordings were typically 3 minutes long. The recording duration was controlled to minimize the effects of removing MEAs from the incubator.

\subsection{Spike Detection}
For each MEA recording, we performed spike detection~\cite{waveclus}. Extracellular signals were band pass filtered using a digital 2nd order Butterworth filter with cutoff frequencies of 0.2 and 4 kHz. Spikes were then detected using a threshold of 5 times the standard deviation of the median noise level. Since there are 120 electrodes in our MultiChannel Systems, the spike detection result of a 3 min recording is a $120 \times 180000$ shape matrix made up of 1s and 0s, where 1 represents neuron firing and 0 represents not firing.

\subsection{Classification}

\begin{figure}\label{Fig_raster}
\includegraphics[width=\textwidth]{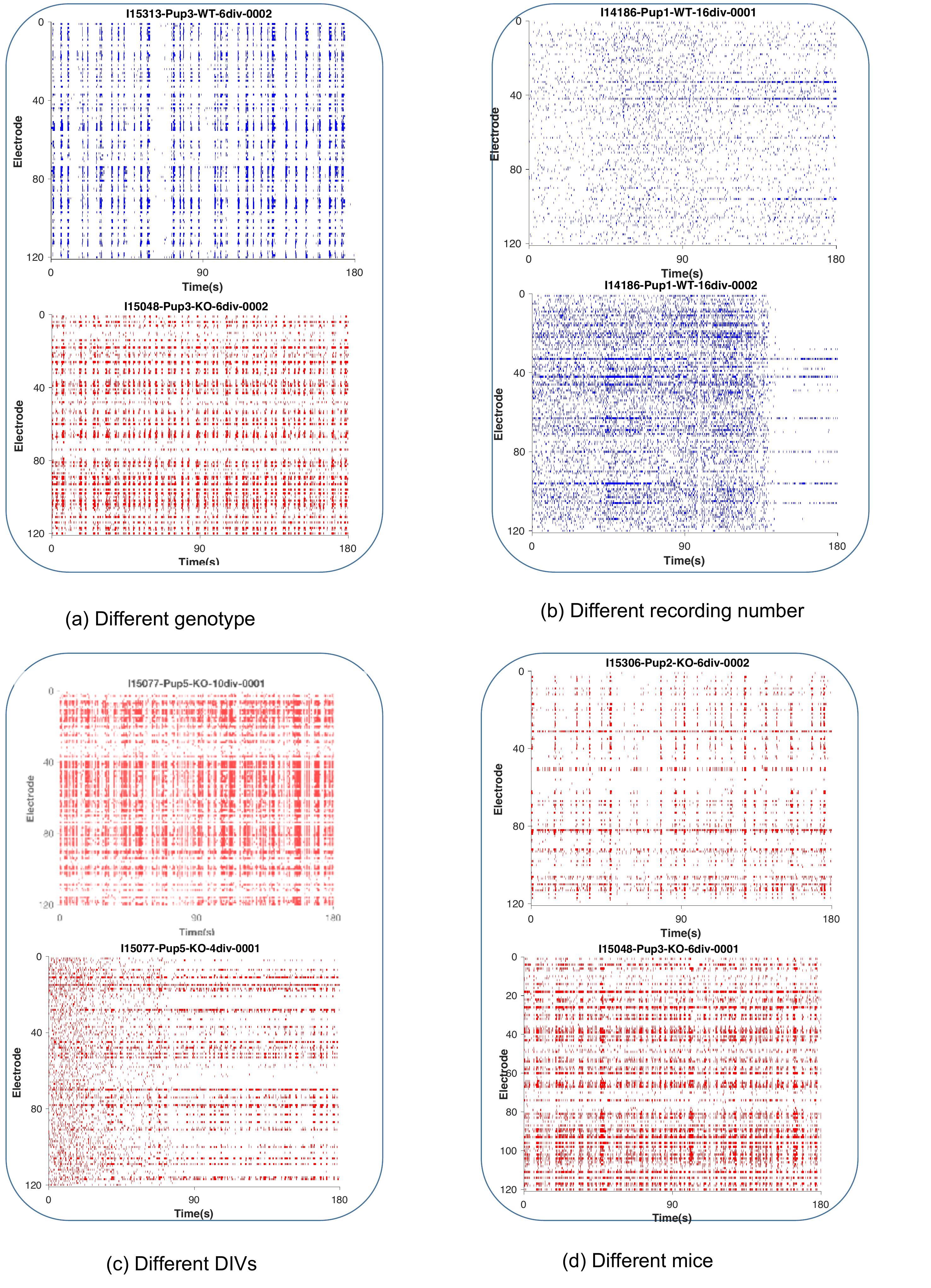}
\caption{Raster Plots of WT and delta-catenin KO. Blue represents WT and red indicates delta-catenin KO. The title of each raster plot is formatted as "MEA device-Mouse-Gene type-DIV-Record Number". (a) KO and WT share some common firing patterns. (b) Different recordings with the same gene type, as well as the same DIV look different. (c) Recordings with the same mouse, the same gene type but different DIV look different. (d) Recordings with the same gene type, the same DIV, but different mouse look different.} \label{fig1}
\end{figure}

For the remainder of this paper, Wild Type (WT) means that there is no gene mutation. Knockout (KO) means that the gene delta-catenin is knocked out or not expressed in the mouse neurons. WS is Williams syndrome neurons, compared with Control. Fig.~\ref{fig1} shows Raster Plots for some sample mouse MEA recordings from WT and KO. From the figure, the recording patterns vary drastically according to different mice, different DIV and even different recording numbers. However, recordings of different genotypes sometimes perform similarly. It is challenging for human eyes to distinguish KO from WT. There are several reasons: 1. The recordings are noisy due to the errors in measuring potentials and spike detection. 2. The firing pattern will change drastically according to many factors like different DIV, different mice and even different recordings. A deep learning classification framework is therefore introduced to automatically predict the genotype, given an MEA recording.

We use two separate sets of MEA recordings in our classification: one dataset consists of mouse neuron recordings to classify KO and WT, while the other dataset consists of human iPSC neuron recordings to distinguish WS and Control human cells. Our mouse recordings consist of 5 separate experiments (Exp1, Exp2, Exp3, Exp4, Exp5) and 331 180000 ms recordings in total, of which 198 recordings are WT and the remainder are delta-catenin KO. Our iPSC recording data are made up of 12 WS recordings and 8 Control recordings. Considering the size of the two datasets, we randomly shuffle and split the mouse MEA data into training, validation and testing by 70\%, 10\% and 20\%, while we apply 5-fold cross-validation for human iPSC recordings.

\section{Deep Learning Model}\label{Model}
\begin{figure}
\includegraphics[width=\textwidth]{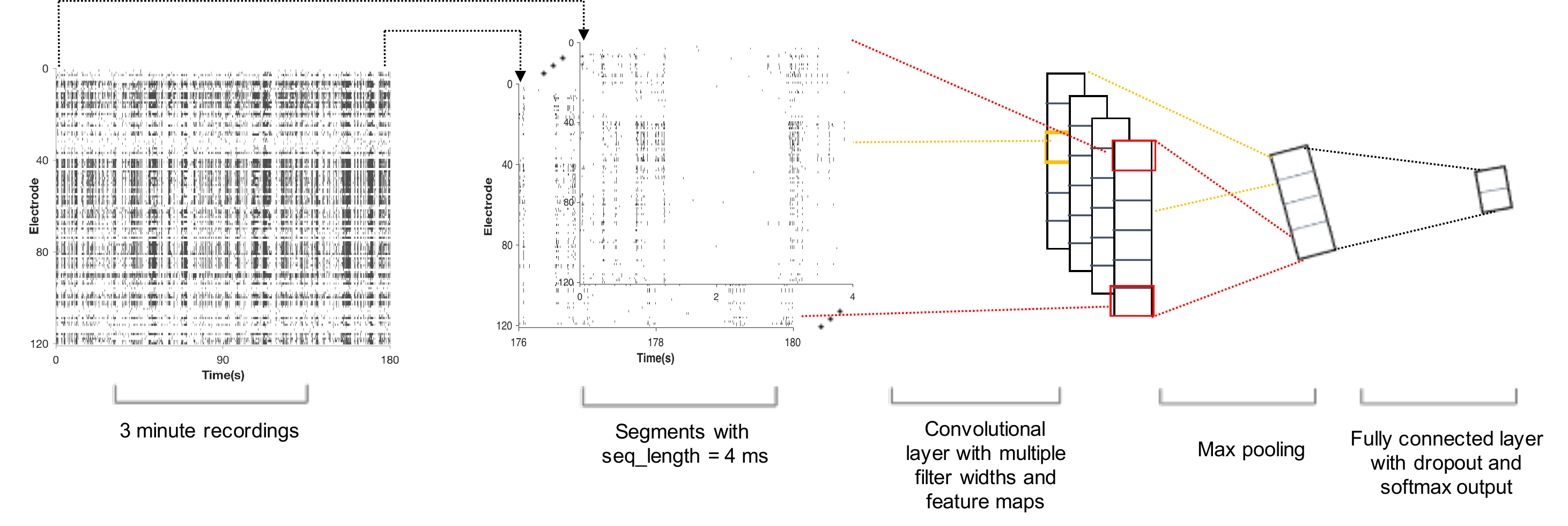}
\caption{Model architecture: 3 minute recordings of the electrical potentials measured on the 120 electrodes are collected form the neuron cultures. Segments with seq\_length = 4 ms of these recordings are individually classified. These individual classifications are conducted for Consensus Prediction in mouse MEA recordings.} \label{Fig_model}
\end{figure}

The model architecture, shown in Fig.~\ref{Fig_model}, consists of convolution-pooling layers followed by fully connected layers.
To learn temporal and spatial invariant features, the convolution is performed on both time and space dimensions. We split the long recordings into smaller slices with length of seq\_length. Detected spikes with shape of (120, seq\_length) serve as input $x$ for the neural network. A convolution operation involves a filter $w \in \mathbb{R}^{st}$, which is applied to a window of s electrodes and t ms to produce a new feature. For example, a feature $f_{i,j},(0 \leq i \leq 120 - s + 1, 0 \leq j \leq seq\_length - t + 1)$ is generated from a window size $(s,t)$ of the spike train:
\begin{equation}
f_{i,j} =ReLU (w · x_{i:i+s-1,j:j+t-1} + b),
\end{equation}
where $b \in \mathbb{R}$ is a bias term. This filter is applied to each possible window of the spike trains to produce a feature map:

\begin{equation}
f=
  \left[ {\begin{array}{cccc}
   f_{1,1} & f_{1,2} & . . . & f_{1,seq\_length-t+1} \\
   f_{2,1} & f_{2,2} & . . . & f_{2,seq\_length-t+1}\\
   . . . & . . . & . . . & . . .\\
   f_{120-s+1,1} & f_{120-s+1,2} & . . . & f_{120-s+1,seq\_length-t+1}
  \end{array} } \right],
\end{equation}
with $f \in \mathbb{R}^{120-s+1,seq\_length-t+1}$. We then apply a max-pooling operation over the feature map and take the maximum value $m = \max{f}$ as the feature corresponding to this particular filter. The idea is to capture the most important feature, the one with the highest value, for each feature map. Our model uses multiple filters to obtain multiple features. These features form the penultimate layer and are passed to a fully connected softmax layer whose output is the probability distribution over two different genotypes. We adjust the number of convolutional ReLU layers from 2 to 5, based on the choice of seq\_length.

We use Batch Normalization~\cite{BatchNorm} to accelerate training. For regulaization, dropout~\cite{Dropout} and early stopping methods~\cite{Earlystop} are implemented to avoid overfitting. Dropout prevents co-adaptation of hidden units by randomly dropping out a proportion of the hidden units during backpropagation. Model training is ended when no improvement is seen during the last 100 validations. Softmax cross entropy loss is minimized with the Adam optimizer~\cite{Adam} for training.

\section{Experimental Setup}\label{Setup}
\subsection{Training and Hyperparameters}
We use 1 ms time bins for our spike train data, thus the dimensionality of time is extremely high. For example, a slice of 10 seconds has 10,000 data points along the time dimension. Thus, the CNN model has a very high demand for memory, while the memory for the graphics processing unit (GPU) is limited. In practice, we randomly sample segments from each recording for training, which not only decreases the GPU memory usage by reducing the dimensionality of time but also increases the number of training samples. For example, if we use seq\_length of 1000ms, then a 180000ms recording can provide 180 independent samples.

We implement the deep learning framework using Tensorflow~\cite{TensorFlow} with the following configurations. The (120, seq\_length) spike detected matrices (see Fig.~\ref{Fig_model}) are input to convolutional ReLU layers which filter the input spike train with $2 \times 5$  kernels and stride of (1, 1). It is interesting to note several biologically inspired hyperparameters in Table~\ref{Bio_tab}. Seq\_length is the slice length that we use to split the recordings. Kernel\_size and stride in CNN correspond to propagation signals, synaptic coupling and correlation between channels. Short latency, monosynaptic, interactions are in a range of 2-4 ms. Propagation signals occurring between nearby electrodes have an average latency of 0.3 ms to 0.7 ms. We choose a kernel size of $2 \times 5$ and stride of (1, 1) to capture propagation signals and synaptic coupling. Hyperparameters are described in Table~\ref{Hyp_tab}. Max pooling is then applied after each convolutional ReLU layer. The feature maps are input for fully connected layers with 2 output nodes for the binary classification.

\subsection{Testing}
For testing, we define Consensus Prediction to measure the performance of predictions for the whole recordings. Consensus Prediction synthesizes results from odd numbers of short slices by majority voting, which can significantly improve the prediction accuracy for a long recording. This is because not all of the short time-slices can be expected to contain useful information. The results of mouse MEA recordings in Section~\ref{Evaluation} are reported with Consensus Prediction.

\subsection{Implementation}
We implement a framework that can distribute the convolutional neural network into multiple ($N$) GPUs to ease the burden on GPU memory. Each GPU contains an entire copy of the deep learning model. We first split the training batch evenly into $N$ sub-batches. Each GPU only processes one of the sub-batches. Then we collect gradients from each replicate of the deep learning model, aggregate them together and update all the replicates. With 3 NVIDIA GeForce GTX 1080s, each of which has a memory of 11178 MB, we can handle spike train segments of 14 seconds with batch size of 24.\\

\begin{table}
\caption{Bio-inspired parameters}\label{Bio_tab}
\begin{center}
\begin{tabular}{|l|l|l|}
\hline
Para &  Biological Justification & value\\
\hline
Seq\_length &  The appropriate slice length which can represent a recording & 4000 ms \\
Kernel\_size &  Propogation signals & (2,5) \\
Stride  & Synaptic coupling, correlation between channels & (1,1)\\

\hline
\end{tabular}
\end{center}
\end{table}

\begin{table}
\caption{Hyperparameters}\label{Hyp_tab}
\begin{center}
\begin{tabular}{|l|l|}
\hline
Hyperparameters  &   Value\\
\hline
Batch size  & 24 \\
Epoch   & 5000\\
Dropout rate   & 0.5\\

\hline
\end{tabular}
\end{center}
\end{table}

\section{Empirical Evaluation}\label{Evaluation}
We focus our evaluation mainly on the accuracy of predicting genotype. We use Consensus Prediction, which is the majority voting result of the sampled short slices, for the mouse recordings. We report the initial prediction accuracy of short slices for human iPSC recordings without Consensus Prediction, since the recording experiments are better controlled.
\subsection{Performance Analysis}
\begin{figure}
\centering
\includegraphics[width=0.8\textwidth]{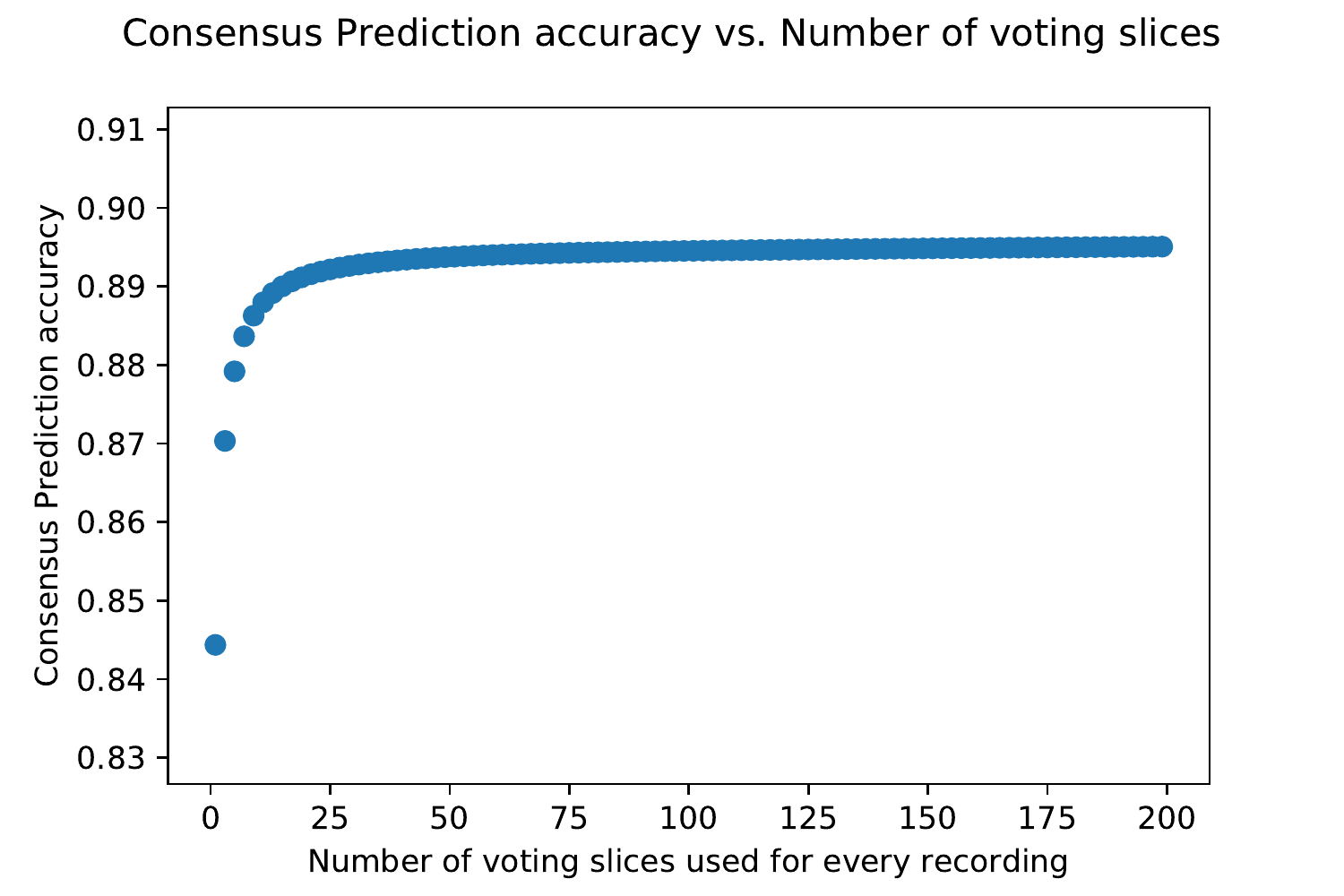}
\caption{Consensus Prediction accuracy vs. number of short slices used for mouse recordings.} \label{whole_accuracy}
\end{figure}

Results of our framework compared against other machine learning models on mouse recordings and human iPSC recordings are shown in Table~\ref{perform_table_1} and Table~\ref{perform_table_2} respectively. We compare our CNN model with Multilayer Perceptron(MLP) and feature based Logistic Regression. We use a two layer MLP, which shares the same hyperparameters with our model's fully connected layers. For Logistic Regression, we first extract features of firing rate and Pearson correlation coefficient between different electrodes for each recording, and then classify neuron genotypes based on these two features. For the mouse recordings, our CNN based deep learning approach improves the Consensus Prediction accuracy by 16.69\% compared with feature based Logistic Regression. Fig~\ref{whole_accuracy} shows the Consensus Prediction accuracy. The accuracy improves by 5.92\% using Consensus Prediction. Although not all of the short slices can be expected to contain enough useful spike patterns, we can overcome that when we synthesize multiple individual classification results from these short slices. For the human iPSC recordings, we report the prediction accuracy of short recording slices. Our model achieves accuracy of 96.18\% even without Consensus Prediction, which is a 15.59\% improvement over feature based Logistic Regression. Our CNN based deep learning model also outperforms MLP on both of the two sets of recordings by 7.00\% and 7.81\% respectively, which shows CNN's advantage of local feature extraction using convolutional kernels over MLP.

\begin{figure}
\centering
\includegraphics[width=0.8\textwidth]{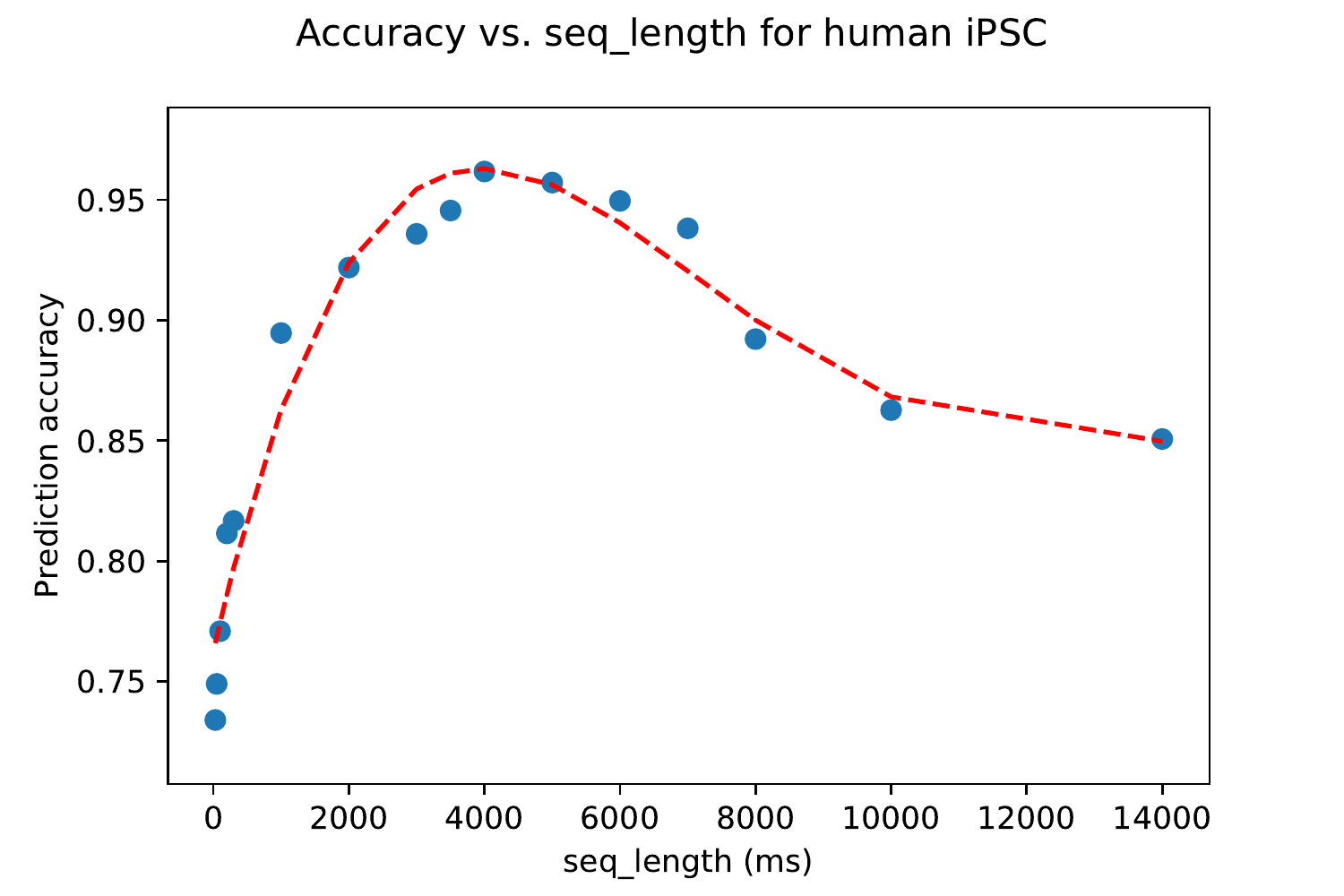}
\caption{Accuracy vs. seq\_length trend for human iPSC recordings.} \label{seq_length}
\end{figure}

Fig~\ref{seq_length} shows the trend of accuracy versus the choice of seq\_length for human iPSC. For the effect of seq\_length on accuracy, there exists a trade off between number of training samples and representation of a whole recording. The short slices contain less information but can provide more independent training samples. For deep learning models, larger numbers of training slices help more than a larger sample. However, we still cannot choose too small of a seq\_length, since a too short slice is not representative for a recording. Given the data we currently have, we use a seq\_length of 4000 ms.

Dropout proved to be such a good regularizer that it was fine to use a larger than necessary network or train too many epochs and simply let dropout regularize it~\cite{Dro}. Dropout consistently added 2\% - 4\% relative performance. Our model converged best with Adam optimizer compared with Vanilla gradient descent, Adagrad~\cite{Adagrad}, Adadelta~\cite{Adadelta} and RMSprop~\cite{RMSprop}.

\begin{table}
\caption{Consensus Prediction performance comparison of our deep learning model with Multilayer Perceptrons and Logistic Regression on mouse recordings.}\label{perform_table_1}
\begin{center}
\begin{tabular}{|l|l|}
\hline
Model &  Accuracy on Testing \\
\hline
\textbf{Convolutional Neural Network} &  \textbf{0.8951} \\
Multilayer Perceptron &  0.8366 \\
Logistic Regression & 0.7671\\

\hline
\end{tabular}
\end{center}
\end{table}

\begin{table}
\caption{Performance comparison of our deep learning model with Multilayer Perceptrons and Logistic Regression on iPSC recordings.}\label{perform_table_2}
\begin{center}
\begin{tabular}{|l|l|}
\hline
Model &  Accuracy on Testing \\
\hline
\textbf{Convolutional Neural Network} &  \textbf{0.9618} \\
Multilayer Perceptron &  0.8921 \\
Logistic Regression & 0.8321\\

\hline
\end{tabular}
\end{center}
\end{table}

\subsection{Case Study}
\begin{figure}
\includegraphics[width=\textwidth]{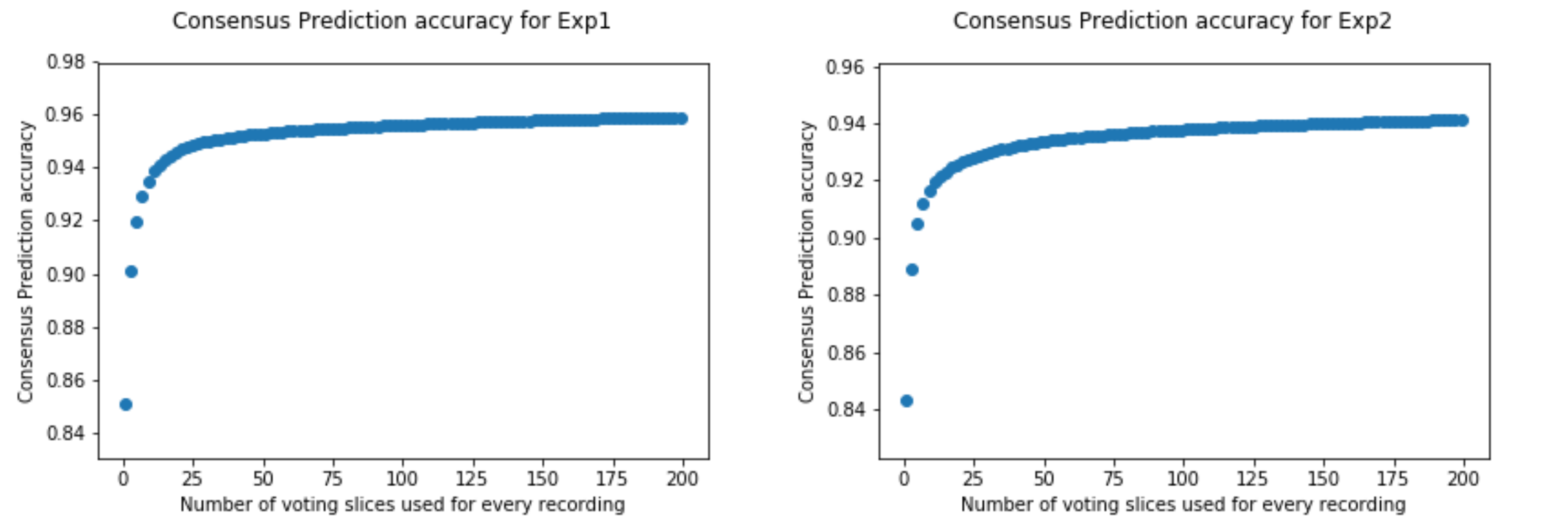}
\caption{Consensus Prediction accuracy for Exp1 and Exp2.} \label{accuracy_Exp5}
\end{figure}

It is challenging to classify the genotype of mouse MEA recordings due to the differences in recordings taken from neurons of different DIV, different mice and different recordings. Considering that the neuron firing patterns change drastically with different DIV, we use two subsets of mouse recording data (Exp1 and Exp2), recorded at 6 DIV and 10 DIV respectively, to study the effect of Consensus Prediction. Fig.~\ref{accuracy_Exp5} shows the prediction accuracy versus number of voting slices in Consensus Prediction. By taking one subset of experiments all recorded at 6 DIV, we achieve a Consensus Prediction accuracy of 95.91\% for Exp1. Similarly, we achieve a Consensus Prediction accuracy of 94.12\% for recordings in Exp2, which are all at 10 DIV. Using Consensus Prediction, we improve the prediction accuracy by 12.70\% and 11.68\% for Exp1 and Exp2 respectively, which indicates that combining information from different parts of one recording significantly helps improve the performance.

\section{Discussion}\label{Discussion}
We have addressed the issue of classifying different genotype MEA recordings by proposing a deep learning framework. We split the long recordings into smaller slices, which not only eases the burden on GPU memory but also provides more training samples for the deep learning model. We use Consensus Prediction during testing, to predict the genotype for a recording. This paper is a proof of principle for classification via deep learning of in-virtro MEA recordings. Clearly, however, more work is needed before it can be known if deep learning will be a generally useful technique for classification of neural cell genotypes or drug effects from in vitro MEA recordings. For example, one can use more recordings and MEAs with larger numbers of probes in future work.

\section{Acknowledgement}
This research was sponsored by the U.S. Army Research Laboratory and Defense Advanced Research Projects Agency under Cooperative Agreement Number W911NF-15-2-0056, Cohen Veterans Biosciences and the Larry L. Hillblom Foundation. We acknowledge helpful comments from Ken Tovar.

%
%
%

\begin{thebibliography}{8}

\bibitem{Imagenet}
Krizhevsky, A., Sutskever, I., and Hinton, G. E.: Imagenet classification with deep convolutional neural networks. In: Advances in Neural Information Processing Systems, pp. 1097-1105. (2012)

\bibitem{Speech}
Graves, A., Mohamed, A. R., and Hinton, G.: Speech recognition with deep recurrent neural networks. In: 2013 IEEE International Conference on Acoustics, Speech and Signal Processing, pp. 6645-6649. (2013)

\bibitem{NLP}
Collobert, R., Weston, J.: A unified architecture for natural language processing: Deep neural networks with multitask learning. In: Proceedings of the 25th international conference on Machine learning, pp. 160–167. ACM (2008)

\bibitem{Alphago}
Silver, D., Schrittwieser, J., Simonyan, K., Antonoglou, I., Huang, A., Guez, A., Hubert, T., Baker, L., Lai, M., Bolton, A. and Chen, Y.: Mastering the game of go without human knowledge. Nature, 550(7676), pp. 354. (2017)

\bibitem{EEG2}
Van Leeuwen, K. G., H. Sun, M. Tabaeizadeh, A. F. Struck, M. J. A. M. Van Putten, and M. B. Westover.: Detecting abnormal electroencephalograms using deep convolutional networks. Clinical Neurophysiology 130 (1), pp. 77-84. (2019)

\bibitem{EEG1}
Schirrmeister, R.T., Springenberg, J.T., Fiederer, L.D.J., Glasstetter, M., Eggensperger, K., Tangermann, M., Hutter, F., Burgard, W. and Ball, T.: Deep learning with convolutional neural networks for EEG decoding and visualization. Human Brain Mapping 38(11), pp. 5391-5420. (2017)


\bibitem{CNN_Neuron}
Kell, A.J., Yamins, D.L., Shook, E.N., Norman-Haignere, S.V. and McDermott, J.H.: A task-optimized neural network replicates human auditory behavior, predicts brain responses, and reveals a cortical processing hierarchy. Neuron, 98(3), pp. 630-644. (2018)

\bibitem{Application}
Buccino, A.P., Ness, T.V., Einevoll, G.T., Cauwenberghs, G. and Häfliger, P.D.: Localizing neuronal somata from multi-electrode array in-vivo recordings using deep learning. In: 39th Annual International Conference of the IEEE Engineering in Medicine and Biology Society (EMBC), pp. 974-977. IEEE (2017)

\bibitem{CNN}
LeCun, Y., Bottou, L., Bengio, Y. and Haffner, P.: Gradient-based learning applied to document recognition. In: Proceedings of the IEEE, 86(11), pp. 2278-2324. (1998)

\bibitem{Simul1}
Buccino, A.P., Kordovan, M., Ness, T.V.B., Merkt, B., Häfliger, P.D., Fyhn, M., Cauwenberghs, G., Rotter, S. and Einevoll, G.T.: Combining biophysical modeling and deep learning for multi-electrode array neuron localization and classification. Journal of neurophysiology. (2018)

\bibitem{Simul2}
Buccino, A.P., Ness, T.V., Einevoll, G.T., Cauwenberghs, G. and Häfliger, P.D.: A Deep Learning Approach for the Classification of Neuronal Cell Types. In: 2018 40th Annual International Conference of the IEEE Engineering in Medicine and Biology Society (EMBC), pp. 999-1002. IEEE (2018)

\bibitem{MEA}
Obien, M.E.J., Deligkaris, K., Bullmann, T., Bakkum, D.J. and Frey, U.: Revealing neuronal function through microelectrode array recordings. Frontiers in neuroscience 8, pp. 423. (2015)

\bibitem{LossD}
Turner, T.N., Sharma, K., Oh, E.C., Liu, Y.P., Collins, R.L., Sosa, M.X., Auer, D.R., Brand, H., Sanders, S.J., Moreno-De-Luca, D. and Pihur, V.: Loss of δ-catenin function in severe autism. Nature 520(7545), pp. 51. (2015)

\bibitem{RequiredD}
Matter, C., Pribadi, M., Liu, X. and Trachtenberg, J.T.: Delta-Catenin is required for the maintenance of neural structure and function in mature cortex in vivo. Neuron 64(3), pp. 320-327. (2009)

\bibitem{JunctionD}
Kosik, K.S., Donahue, C.P., Israely, I., Liu, X. and Ochiishi, T.: Delta-Catenin at the synaptic–adherens junction. Trends in cell biology, 15(3), pp. 172-178. (2015)

\bibitem{Williams}
Lalli, M.A., Jang, J., Park, J.H.C., Wang, Y., Guzman, E., Zhou, H., Audouard, M., Bridges, D., Tovar, K.R., Papuc, S.M. and Tutulan-Cunita, A.C.: Haploinsufficiency of BAZ1B contributes to Williams syndrome through transcriptional dysregulation of neurodevelopmental pathways. Human molecular genetics 25(7), pp. 1294-1306. (2016)

\bibitem{Williams1}
Bayés, M., Magano, L.F., Rivera, N., Flores, R. and Jurado, L.A.P.: Mutational mechanisms of Williams-Beuren syndrome deletions. The American Journal of Human Genetics 73(1), pp. 131-151. (2003)

\bibitem{Williams2}
Morris, C.A., Lenhoff, H.M. and Wang, P.P. eds.: Williams-Beuren syndrome: Research, evaluation, and treatment. JHU Press. (2016)

\bibitem{Cut}
Tovar, K.R. and Westbrook, G.L.: Amino-terminal ligands prolong NMDA receptor-mediated EPSCs. Journal of Neuroscience 32(23), pp. 8065-8073. (2012)

\bibitem{Propogation}
Tovar, K.R., Bridges, D.C., Wu, B., Randall, C., Audouard, M., Jang, J., Hansma, P.K. and Kosik, K.S.: Recording action potential propagation in single axons using multi-electrode arrays. bioRxiv, pp. 126425. (2017)

\bibitem{waveclus}
Quiroga, R.Q., Nadasdy, Z. and Ben-Shaul, Y.: Unsupervised spike detection and sorting with wavelets and superparamagnetic clustering. Neural Computation 16(8), pp. 1661-1687. (2014)

\bibitem{TensorFlow}
Abadi, M., Agarwal, A., Barham, P., Brevdo, E., Chen, Z., Citro, C., Corrado, G.S., Davis, A., Dean, J., Devin, M. and Ghemawat, S.: TensorFlow: Large-scale machine learning on heterogeneous systems, 2015. Software available from tensorflow. org, 1(2). (2015)

\bibitem{Dropout}
Srivastava, N., Hinton, G., Krizhevsky, A., Sutskever, I. and Salakhutdinov, R.: Dropout: a simple way to prevent neural networks from overfitting. The Journal of Machine Learning Research 15(1), pp. 1929-1958. (2014)

\bibitem{Earlystop}
Prechelt, L.: Early stopping-but when?. In: Neural Networks: Tricks of the trade, pp. 55-69. Springer, Berlin, Heidelberg. (1998)

\bibitem{BatchNorm}
Ioffe, S. and Szegedy, C.: Batch normalization: Accelerating deep network training by reducing internal covariate shift. arXiv preprint arXiv:1502.03167. (2015)

\bibitem{Adam}
Kingma, D.P. and Ba, J.: Adam: A method for stochastic optimization. arXiv preprint arXiv:1412.6980. (2014)

\bibitem{Dro}
Wu, H. and Gu, X.: Towards dropout training for convolutional neural networks. Neural Networks 71, pp. 1-10. (2015)


\bibitem{Adagrad}
Duchi, J., Hazan, E. and Singer, Y.: Adaptive subgradient methods for online learning and stochastic optimization. Journal of Machine Learning Research, pp.2121-2159. (2011)

\bibitem{Adadelta}
Zeiler, M.D.: ADADELTA: an adaptive learning rate method. arXiv preprint arXiv:1212.5701. (2012)

\bibitem{RMSprop}
Tieleman, T. and Hinton, G.: Lecture 6.5-RMSProp, COURSERA: Neural networks for machine learning. University of Toronto, Technical Report. (2012)

\end{thebibliography}
%

\end{document}